\documentclass[journal=cmatex]{achemso}
\usepackage{xcolor}
\usepackage{graphicx}
\usepackage{geometry}
\usepackage{gensymb}
\usepackage{cleveref}
\usepackage{bm}
\usepackage[version=4]{mhchem}
\usepackage{multirow}
\usepackage{epstopdf}
\usepackage{pdfpages}

\author{Pinwen Guan}
\affiliation{Department of Mechanical Engineering, Carnegie Mellon University, Pittsburgh, Pennsylvania 15213, USA}
\author{Gregory Houchins}
\affiliation[Scott Institute]
{Wilton E. Scott Institute for Energy Innovation, Carnegie Mellon University, Pittsburgh, Pennsylvania 15213, USA}
\alsoaffiliation[Department of Physics]
{Department of Physics, Carnegie Mellon University, Pittsburgh, Pennsylvania 15213, US}
\author{Venkatasubramanian Viswanathan}
\affiliation{Department of Mechanical Engineering, Carnegie Mellon University, Pittsburgh, Pennsylvania 15213, USA}
\email{venkvis@cmu.edu}
\affiliation[Scott Institute]
{Wilton E. Scott Institute for Energy Innovation, Carnegie Mellon University, Pittsburgh, Pennsylvania 15213, USA}
\alsoaffiliation[Department of Physics]
{Department of Physics, Carnegie Mellon University, Pittsburgh, Pennsylvania 15213, US}

\title[Depye]
  {Uncertainty Quantification of DFT-predicted Finite Temperature Thermodynamic Properties within the Debye Model}

\begin{document}
\onehalfspacing
\begin{abstract}

Density functional theory (DFT) calculations are routinely used to screen for functional materials for a variety of applications.  This screening is often carried out with a few descriptors, which uses ground-state properties that typically ignores finite temperature effects.  Finite-temperature effects can be included by calculating the vibrations properties and this can greatly improve the fidelity of computational screening.  An important challenge for DFT-based screening is the sensitivity of the predictions to the choice of the exchange correlation function.  In this work, we rigorously explore the sensitivity of finite temperature thermodynamic properties to the choice of the exchange correlation functional using the built-in error estimation capabilities within the Bayesian Error Estimation Functional (BEEF-vdw).  The vibrational properties are estimated using the Debye model and we quantify the uncertainty associated with finite-temperature properties for a diverse collection of materials.  We find good agreement with experiment and small spread in predictions over different exchange correlation functionals for Mg, \ce{Al2O3}, Al, Ca, and GaAs. In the case of Li, \ce{Li2O}, and NiO, however, we find a large spread in predictions as well as disagreement between experiment and functionals due to complex bonding environments. While the energetics generated by BEEF-vdW ensemble is typically normal, the complex mapping through the Debye model leads to the derived finite temperature properties having non-Gaussian behavior.  We test a wide variety of probability distributions that best represent the finite temperature distribution and find that properties such as specific heat, Gibbs free energy, entropy, and the thermal expansion coefficient are well described by normal or transformed normal distributions, while the prediction spread of volume at a given temperature does not appear to be drawn from a single distribution. Given the computational efficiency of the approach, we believe that uncertainty quantification should be routinely incorporated into finite-temperature predictions. In order to facilitate this, we have open-sourced the code base, under the name, Depye.
\end{abstract}
\section{Introduction}
Ab initio calculations utilizing density functional theory have enabled high throughput materials discovery for a variety of applications such as batteries \cite{Kang2006}, thermoelectrics\cite{Zhang2016}, solar photovoltaics \cite{yu2012identification}, solar fuels \cite{castelli2012computational}. These high throughput searches typically involve calculating a few ground state properties (descriptors) that often correlate well with the required output performance.  However, in most cases, the materials need to perform at finite temperature, and typically the finite temperature effects are neglected.  Explicit calculation of vibrational properties through density functional perturbation theory \cite{Baroni2001}, or with finite difference \cite{kresse1995} is too computationally intense for most applications and makes it difficult to incorporate in materials screening studies. 

The Debye-Gr{\"u}neisen model has been used successfully to understand thermodynamic properties of solids by treating the materials as an elastic continuum with linear phonon dispersion. \cite{Moruzzi1988,LU20071215} While rather simple, this approximation has worked satisfactorily for a large range of materials including metals \cite{Moruzzi1988}, carbides \cite{LU20071215}, nitrides \cite{LU20071215}, sulfides \cite{Cui_E_2008}, oxides \cite{BLANCO1996245} and fluorides\cite{BLANCO1996245}. The reason for this success is that in many cases only an average of particular phonon frequencies is needed to calculate many thermophysical properties. \cite{GRIMVALL1999112} While the greatly simplified phonon spectrum used within the Debye model is sufficient for many materials, is is not valid for materials with high inhomogeneity in structural and chemical bonding which leads to a large separation in the low-frequency and high-frequncy regimes of the phonon spectrum \cite{GUAN2017510}. This simplistic and computationally efficient method can be improved further with the use of a scaling factor to account for varying speeds of sound in the transverse and longitudinal directions \cite{CHEN2001947}.  The use of the Debye-Gr{\"u}neisen model allows for the approximation of the vibrational properties with a fraction of the computational cost of explicit phonon computations while still harnessing the flexibility and first principles nature of density functional theory. 

At the same time, a major source of error within density functional theory calculations is the choice of exchange correlation functional.\cite{BEEF} While there exists rules of thumb for picking suitable exchange correlation functionals given the type of system being studied, currently, no functional can equally well describe all material systems, nor is there much quantitative understanding of the error in the prediction for an arbitrary material with no prior experimental knowledge.  The sensitivity of prediction results with respect to exchange correlation functional typically remains uncertain. This effect is compounded for complex properties that involve derivatives of the energy where errors can propagate, for e.g. thermodynamic predictions using the Debye model with DFT input data. One approach to address this has been to use Bayesian statistics to provide an ensemble of functionals that can be evaluated non-self-consistently, and therefore computationally efficiently\cite{BEEF}. The spread in the resulting DFT estimation of energy has been trained to match the error of the self consistent results with respect to experiment. Additionally, it has be demonstrated that this spread of energies bounds other general gradient approximation results for magnetic properties \cite{PhysRevB.96.134426}, elastic properties\cite{PhysRevB.94.064105}, surface phase diagrams\cite{acs.langmuir.8b02219,acscatal.8b01432}, scaling relations in oxygen reduction\cite{Yan2017,Deshpande2016}, formation enthalpies of reactions\cite{Rune2015CST}, vibrational frequencies \cite{doi:10.1021/acs.jpcc.8b11689} and energetics of electrochemical reactions.\cite{acs.jpclett.7b02895}. 

In this work, we describe a computationally efficient approach that propagates the error estimation capabilities built within BEEF-vdW exchange correlation functional through the Debye Model to derive uncertainty associated with the predicted thermodynamic properties.  We show that the distribution obtained bounds other GGA-level functionals for a wide range of materials and bonding environments.  Given the complex nature of the model, we perform rigorous statistical analysis to determine the nature of the output distribution for the predicted thermodynamic properties through goodness of fit tests.  We believe that the derived uncertainty can be used in conjuction with design of experiments (DOE) approaches that perform a hybrid of explore and exploit, which performs much better than a simple exploit approach for complex non-linear material design problems.\cite{ling2017high}  Given the ease of integration and computational efficiency of the present approach, we believe that this can be routinely incorporated into high-throughput material searches.

\section{Methods}
\subsection{Calculation Details}
All calculation were run using a real space grid implementation of the projector augmented wave approximation through the GPAW software \cite{PhysRevB.71.035109,Enkovaara_2010,ase-paper}. A real space grid with spacing of 0.16 \AA ~ is used for the representation of electronic wavefunctions, and a k-point density of larger than 30 \AA~  in reciprocal space was used in each dimension. For each material, the geometry is relaxed to a maximum force of less than 0.01 eV/\AA.

\subsection{Debye Model}

\begin{figure}[H]
\includegraphics[width=\textwidth]{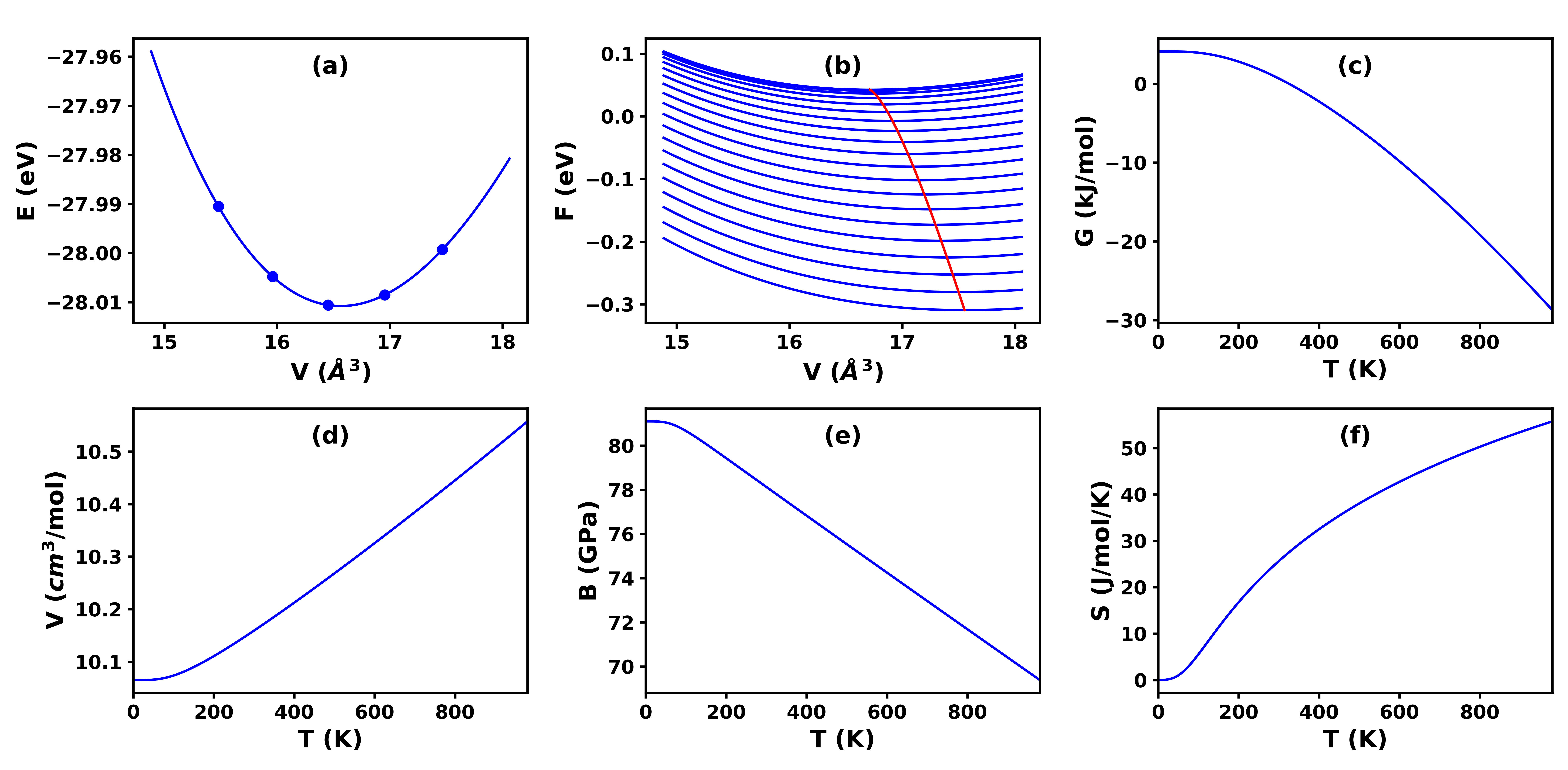}
\centering
\caption{Procedure of calculating thermodynamic properties based on the Debye model: (a) E-V curve (b) free energies at different temperatures based on the Debye model with the red line consisting of the minimum of free energy at each temperature, from which (c) Gibbs energy (d) volume (e) bulk modulus and (f) entropy are derived.} 
\label{fig:thermo}
\end{figure}

 The energies of 5 volumes around the minimum is computed. From these energy-volume values, an Vinet equation of state was fit \cite{PhysRevB.35.1945} providing the miniumium energy, minimium volume $V_0$, and bulk modulus $B_0$ at zero temperature. For BEEF containing thousands of functionals with the equilibrium volumes span a considerable range, a total of 16 volumes are computed, among which the 5 most energetically favorable volumes are used for fitting for the corresponding functional. The Debye temperature is then calculated as 

\[ \Theta_D = sAV_{0}^{1/6} \left(\frac{B_0}{M}\right)^{1/2}  \left(\frac{V_0}{V}\right)^{\gamma}    \] 

\noindent where constant $ A=(6\pi^2)^{1/3}\frac{\hbar}{k_{B}} $, and s is the scaling factor. Moruzzi et al. recommended using a scaling factor of $ s=0.617 $, based on a study of 14 nonmagnetic cubic metals \cite{Moruzzi1988}. Alternatively, the scaling factor can be derived as in ref. \citenum{CHEN2001947}. 

\[ s(\nu)=3^{5/6}\left[ 4 \sqrt{2} \left(\frac{1+\nu}{1-2\nu}\right)^{3/2} + \left(\frac{1+\nu}{1-\nu}\right)^{3/2} \right]^{-1/3}   \] 
where $\nu$ is the Poisson ratio. A more detailed discussion about the scaling factor used in the present work is given in the Supporting Information. The Gr{\"u}neisen parameter $\gamma$ can be approximated as $\gamma=\frac{1+B_0^\prime}{2}-x$, where $ B_0^\prime $ is the derivative of the bulk modulus with respect to pressure and is dimensionless. The term x is set to 1 to best describe properties below the Debye temperature and $ \frac{2}{3} $ for properties above \cite{Moruzzi1988}. In the present work $ x=\frac{2}{3} $ is used to study the thermodynamic properties at elevated temperatures.

The vibrational free energy can then be calculated as 

\[F_{vib}=\frac{9}{8} k_{B} \Theta_D (V) - k_{B}T \left( D\left[\frac{\Theta_D(V)}{T} \right]+ 3 \textrm{ln}[1-\textrm{e}^{-\Theta_D(V)/T}]  \right) \]

\noindent where the Debye function is defined as $D(x)=\frac{3}{x^3}\int_0^x \frac{z^3 \textrm{dz}}{\textrm{e}^z-1}$. Additionally we compute the vibrational isochoric heat capacity $C_V$ as:

\[C_v = 9Nk_B \left(\dfrac{T}{T_D}\right)^3 \int_0^{T_D/T} \dfrac{x^4 \mathrm{e}^{x}}{(\mathrm{e}^{x}-1)^2} \mathrm{d}x\]

We ignore contributions from the other sources, which are typically less important, e.g., the thermal excitation of electrons \cite{RevModPhys.74.11}.  The free energy is $F(V,T)=E(V)+F_{vib}(V,T) $ where $E$ is the static energy.

At ambient conditions, the pressure effect on the Gibbs Free Energy is small as shown in the Supporting Information. Thus, the pressure effect can be ignored and it is valid to use Gibbs Free Energy at zero pressure

\[ G(P=0,T)=\min\limits_{V} F(V,T) \]

with the corresponding volume $ V_0(T) $ and bulk modulus $ B_0(T) $.

For simplicity, $ G(P=0,T) $ is written as $ G $ hereafter. The entropy is then derived as, $S=-\frac{dG}{dT}$ and the enthalpy as $H=G+TS$
The isobaric heat capacity can be obtained by, $C_p=\frac{dH}{dT}$. Finally, the volumetric thermal expansion coefficient is calculated as $\alpha=\frac{1}{V}\frac{dV_0}{dT}$. All derivatives with respect to T were taking using the central difference approximation.

\subsection{Bayesian Error Estimation Funtional}
For each material, a collection of functionals at the level of the generalized gradient approximation (GGA) were used as described below. For the purpose of error estimation, the Bayesian error estimation functional with van der Waals correction \cite{BEEF} was used. Besides acting in the usual manner for the self consistent calculation of the exchange correlation energy, this empirically fit functional can generate an ensemble of functionals that are small perturbations away from the main functional fit in exchange correlation space.

The exchange-correlation energy for the BEEF-vdW is given by
	\[E_{xc}=\sum_{m}\int \epsilon_x^{UEG}(n)B_m[t(s)]d\bm{r} 
	+\alpha_c E^{LDA-c}+(1-\alpha_c)E^{PBE-c}+E^{nl-c}\]

The method to generate the ensemble of functionals was tuned such that the spread of the predictions of the functionals matches the error of the main self consistent functional with respect to the training and experimental data. Each of these functionals can then provide a non self consistent prediction of energy and therefore allows for a computationally efficient yet systematic way of understanding the sensitivity of the final prediction with respect to small changes in exchange correlation functional. 

A conventional way of uncertainty propagation would be to take the distribution generated from an DFT error estimation technique like BEEF, or assume distribution for each of the inputs into the thermodynamic model, and then propagate that error forward in some way.\cite{PAULSON201974} This propagation of uncertainty would treat the uncertainty as if the errors of each of the inputs is independent and ultimately lead to a slight overestimation of the resulting error attributed to exchange correlation functional uncertainty. This is because the error for one property predicted by a function is not statistically independent from the error for another property predicted by the same functional. By propagating the ensemble of energetic predictions of each material at each volume, we can generate an ensemble of thermodynamic predictions and therefore an better estimation of uncertainty in the predictions due to the underlying DFT data. 

\section{Results and discussion}

\begin{table}
\begin{center}
\footnotesize
\begin{tabular}{ |p{0.7cm}|p{0.7cm}|p{1.1cm}|p{1.1cm}|p{1.2cm}|p{1.1cm}|p{1.1cm}|p{1.1cm}|p{1.1cm}|p{1.1cm}|c|c|c| } 
\hline
 & Prop. & PBE & RPBE & optPBE-vdW  &  PBEsol &  PW91 & BEEF-vdW & mu & sigma & COV & Skew & Kurt \\
\hline
 \multirow{6}{4em}{Al} 
  & G & 0.46 & 0.41 & 0.03 & 0.73 & 0.35 & 0.67 & 0.58 & 0.87 &  & -0.98 & 0.99   \\
  & S & 26.35 & 26.54 & 27.78 & 25.63 & 26.76 & 25.63 & 26.09 & 2.73 & 0.1 & 1.14 & 1.34   \\
  & Cp & 23.6 & 23.69 & 24.25 & 23.45 & 23.83 & 23.19 & 23.54 & 1.12 & 0.05 & 1.47 & 2.5   \\
  & BT & 73.69 & 72.05 & 64.69 & 78.94 & 71.04 & 78.13 & 77.56 & 17.23 & 0.22 & -0.39 & -0.23   \\
  & VT & 10.16 & 10.38 & 10.51 & 10.0 & 10.26 & 10.16 & 10.22 & 0.68 & 0.07 & 0.57 & 0.85   \\
  & TEC & 62.65 & 64.86 & 79.39 & 60.59 & 68.92 & 50.64 & 62.16 & 28.97 & 0.47 & 1.74 & 3.6   \\
 \hline
 \multirow{6}{4em}{Ca} 
  & G & -4.78 & -5.01 & -4.72 & -4.7 & -4.79 & -4.75 & -4.57 & 0.78 &  & -0.37 & 0.32   \\
  & S & 41.97 & 42.76 & 41.84 & 41.71 & 42.02 & 41.78 & 41.37 & 2.59 & 0.06 & 0.47 & 0.33   \\
  & Cp & 25.16 & 25.31 & 25.24 & 25.18 & 25.23 & 25.0 & 25.22 & 1.05 & 0.04 & 1.54 & 2.45   \\
  & BT & 16.31 & 15.0 & 16.54 & 16.82 & 16.25 & 16.39 & 17.36 & 3.61 & 0.21 & 0.38 & 0.66   \\
  & VT & 26.08 & 27.77 & 25.91 & 25.36 & 26.11 & 26.69 & 26.62 & 2.11 & 0.08 & -0.36 & -0.43   \\
  & TEC & 81.1 & 87.13 & 84.79 & 82.29 & 84.43 & 72.14 & 70.46 & 52.98 & 0.75 & 0.41 & 0.51   \\
 \hline
 \multirow{6}{4em}{Li} 
  & G & -0.89 & -0.81 & -0.86 & -0.88 & -0.85 & -0.39 & -0.09 & 0.97 &  & -0.23 & 0.34 \\
  & S & 31.02 & 30.65 & 30.88 & 30.9 & 30.78 & 29.59 & 28.45 & 2.98 & 0.1 & 0.47 & 0.41 \\
  & Cp & 25.91 & 25.56 & 25.75 & 25.71 & 25.61 & 25.71 & 24.87 & 1.69 & 0.07 & 0.64 & 0.24 \\
  & BT & 11.64 & 11.82 & 11.7 & 11.73 & 11.82 & 13.31 & 14.95 & 3.89 & 0.26 & 0.52 & 0.61 \\
  & VT & 13.21 & 13.62 & 13.38 & 13.18 & 13.23 & 12.7 & 12.73 & 1.31 & 0.1 & 0.41 & -0.03 \\
  & TEC & 232.27 & 212.58 & 223.4 & 222.49 & 217.45 & 220.87 & 167.05 & 96.26 & 0.58 & -0.03 & 0.3 \\
 \hline
 \multirow{6}{4em}{Mg} 
  & G & -2.67 & -2.76 & -2.78 & -2.44 & -2.66 & -2.55 & -2.52 & 0.56 &  & -0.32 & 0.71 \\
  & S & 35.63 & 35.87 & 35.97 & 34.82 & 35.55 & 35.2 & 35.15 & 1.79 & 0.05 & 0.59 & 1.25 \\
  & Cp & 25.17 & 25.09 & 25.19 & 24.86 & 25.05 & 24.98 & 25.01 & 0.64 & 0.03 & 1.37 & 3.79 \\
  & BT & 33.47 & 32.44 & 32.36 & 35.94 & 33.64 & 34.5 & 35.08 & 5.34 & 0.15 & 0.17 & 0.83 \\
  & VT & 14.2 & 14.63 & 14.41 & 13.91 & 14.2 & 14.33 & 14.36 & 0.72 & 0.05 & 0.22 & 0.38 \\
  & TEC & 94.35 & 90.39 & 94.81 & 82.71 & 89.58 & 86.57 & 86.69 & 25.04 & 0.29 & 0.9 & 2.68 \\
 
 \hline
 
 \multirow{6}{4em}{NiO} 
  & G & 6.06 & 4.75 & 4.62 & 6.39 & 5.46 & 4.69 & 4.89 & 1.52 &  & -0.18 & 0.17   \\
  & S & 39.27 & 42.4 & 42.81 & 38.55 & 40.6 & 42.66 & 42.28 & 3.92 & 0.09 & 0.4 & 0.5   \\
  & Cp & 43.34 & 44.24 & 44.5 & 43.13 & 43.57 & 44.52 & 44.36 & 1.97 & 0.04 & 0.47 & 2.23   \\
  & BT & 166.19 & 142.01 & 140.11 & 177.3 & 156.86 & 140.85 & 145.88 & 27.25 & 0.19 & 0.33 & 0.17   \\
  & VT & 12.06 & 12.36 & 12.23 & 11.07 & 11.77 & 12.31 & 12.32 & 0.62 & 0.05 & 0.18 & 0.16   \\
  & TEC & 45.23 & 47.17 & 50.67 & 46.68 & 43.57 & 51.39 & 45.67 & 27.27 & 0.6 & 0.13 & 0.68   \\
 \hline
 \multirow{6}{4em}{GaAs} 
   & G & -3.77 & -4.21 & -3.91 & -2.87 & -3.58 & -3.88 & -1.87 & 0.38 &  & 0.11 & 0.72   \\
  & S & 66.17 & 67.46 & 66.49 & 63.23 & 65.44 & 66.43 & 33.01 & 1.16 & 0.04 & 0.1 & 0.88   \\
  & Cp & 49.09 & 49.22 & 48.95 & 48.27 & 48.72 & 48.99 & 24.48 & 0.34 & 0.01 & 1.03 & 2.29   \\
  & BT & 57.27 & 53.66 & 56.05 & 65.58 & 59.0 & 56.32 & 57.68 & 6.38 & 0.11 & 0.52 & 1.5   \\
  & VT & 29.39 & 30.35 & 30.02 & 28.13 & 29.31 & 29.84 & 14.91 & 0.78 & 0.05 & 0.09 & -0.13   \\
  & TEC & 59.2 & 60.15 & 56.13 & 47.93 & 53.27 & 56.96 & 56.32 & 11.18 & 0.2 & 0.55 & 1.24   \\
 \hline
 \multirow{6}{4em}{\ce{Li2O}}
  & G & 22.34 & 19.0 & 20.28 & 20.99 & 20.36 & 21.54 & 22.9 & 3.81 &  & -0.09 & 0.35   \\
  & S & 33.74 & 40.15 & 37.87 & 36.46 & 37.59 & 35.21 & 33.3 & 6.44 & 0.19 & 0.56 & 0.81   \\
  & Cp & 52.93 & 58.49 & 56.84 & 55.53 & 56.44 & 54.27 & 52.19 & 5.96 & 0.11 & 0.24 & -0.1   \\
  & BT & 72.39 & 55.33 & 61.08 & 65.69 & 62.39 & 66.87 & 74.7 & 19.16 & 0.26 & 0.57 & 0.5   \\
  & VT & 15.66 & 17.57 & 16.93 & 15.84 & 16.27 & 16.85 & 16.69 & 1.06 & 0.06 & 0.01 & -0.56   \\
  & TEC & 79.86 & 110.44 & 103.05 & 96.65 & 100.68 & 85.39 & 69.82 & 45.32 & 0.65 & 0.29 & 0.46   \\
 \hline
 \multirow{6}{4em}{\ce{Al2O3}} 
  & G & 41.6 & 38.36 & 39.81 & 41.33 & 40.28 & 39.1 & 39.29 & 2.37 &  & 0.14 & -0.31 \\
  & S & 47.43 & 52.4 & 50.17 & 47.89 & 49.47 & 51.32 & 51.12 & 3.74 & 0.07 & 0.11 & -0.15 \\
  & Cp & 78.77 & 83.18 & 81.28 & 79.24 & 80.68 & 82.31 & 82.06 & 3.4 & 0.04 & 0.17 & 0.11 \\
  & BT & 223.54 & 194.8 & 206.81 & 221.08 & 211.34 & 201.05 & 203.4 & 20.33 & 0.1 & 0.33 & -0.22 \\
  & VT & 26.77 & 28.16 & 27.7 & 26.85 & 27.36 & 27.81 & 27.74 & 1.01 & 0.04 & -0.17 & -0.49 \\
  & TEC & 24.04 & 29.3 & 27.26 & 25.11 & 26.79 & 28.85 & 28.15 & 7.46 & 0.27 & 0.64 & 1.71 \\
 \hline

\end{tabular}
\caption{\label{tab:stat1}Properties and their statistics at 300 K.}
\end{center}
\end{table}

\begin{figure}[H]
\includegraphics[width=\textwidth]{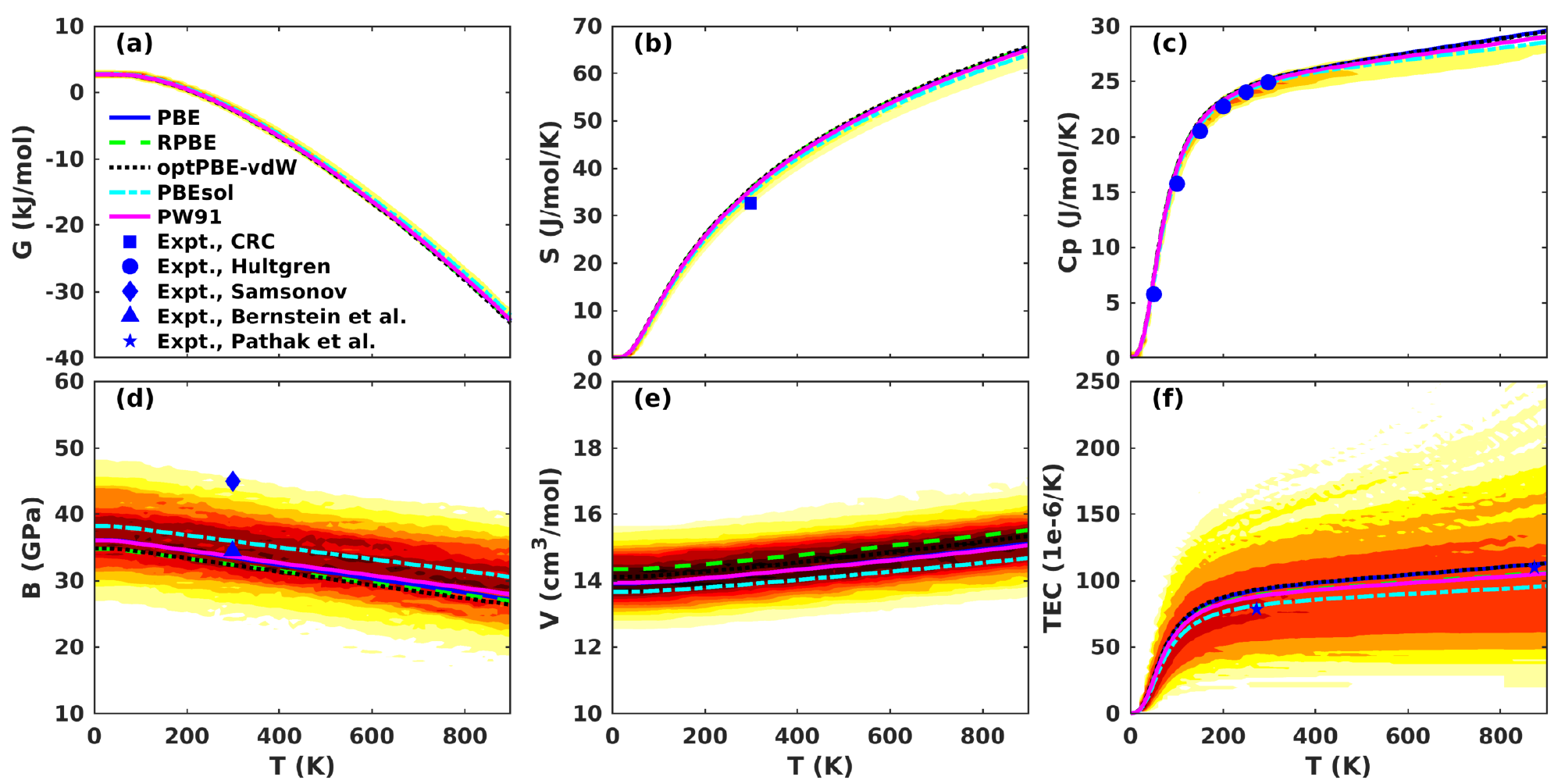}
\centering
\caption{Thermodynamic properties of hcp Mg: (a) Gibbs energy (b) entropy (c) isobaric heat capacity (d) bulk modulus (e) volume and (f) volumetric thermal expansion coefficient from DFT Debye calculations compared with experimental data in the literature \cite{alma991000880309704436,hultgren1963selected,samsonov2012handbook,663214302,doi:10.1002/pssa.2210660271}. The colormap represents the probability distribution function (PDF) of each property calculated by BEEF. The natural logarithm of PDF is presented in (f) for better view.}
\label{fig:Mg}
\end{figure}

\begin{figure}[h]
\includegraphics[width=\textwidth]{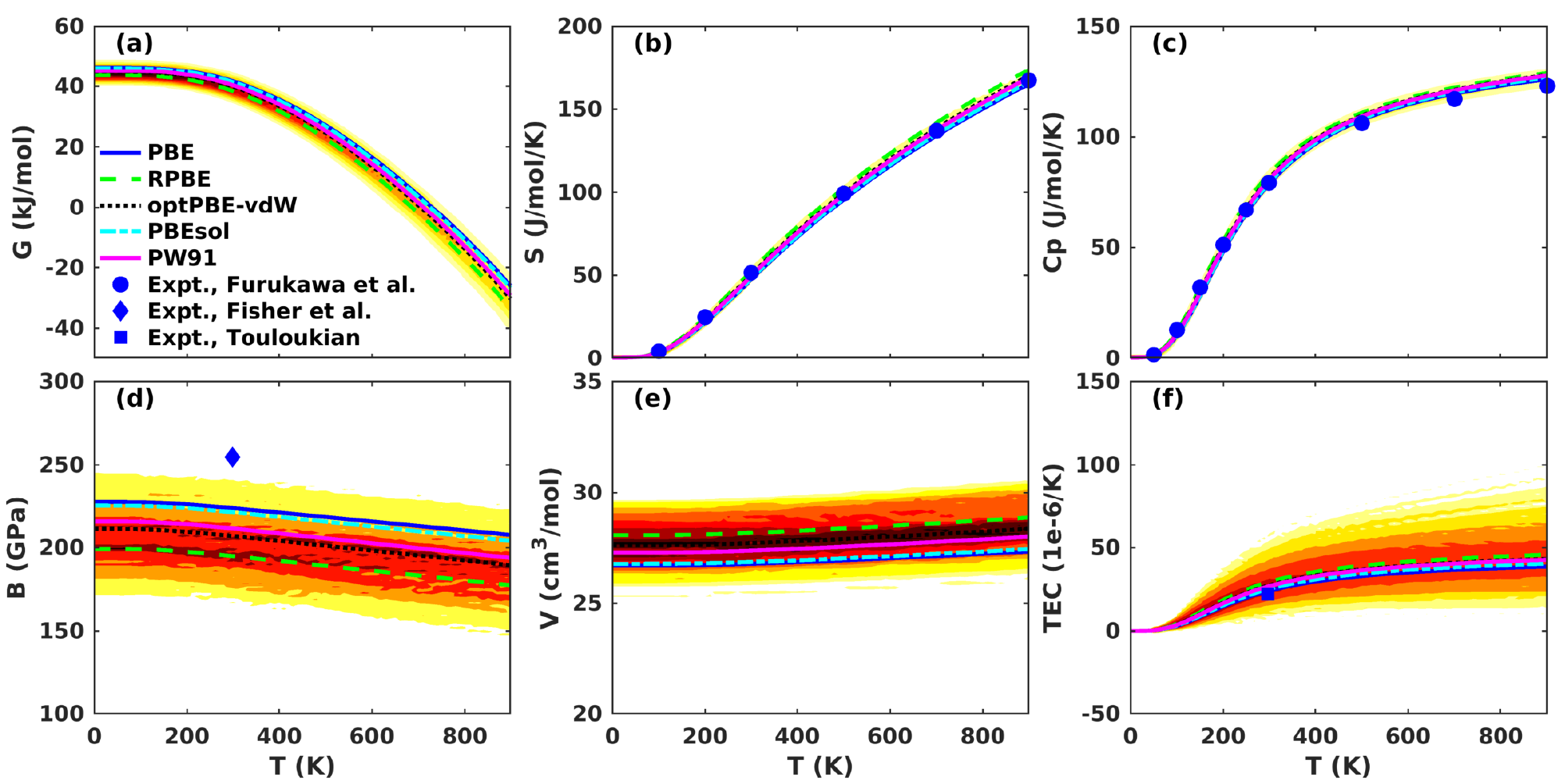}
\centering
\caption{Thermodynamic properties of corundum \ce{Al2O3}: (a) Gibbs energy (b) entropy (c) isobaric heat capacity (d) bulk modulus (e) volume and (f) volumetric thermal expansion coefficient from DFT Debye calculations compared with experimental data in the literature \cite{10015561638,doi:10.1111/j.1151-2916.1992.tb04159.x,touloukian1967thermophysical}. The colormap represents the PDF of each property calculated by BEEF. The natural logarithm of PDF is presented in (f) for better view.}
\label{fig:Al2O3}
\end{figure}

\begin{figure}[h]
\includegraphics[width=\textwidth]{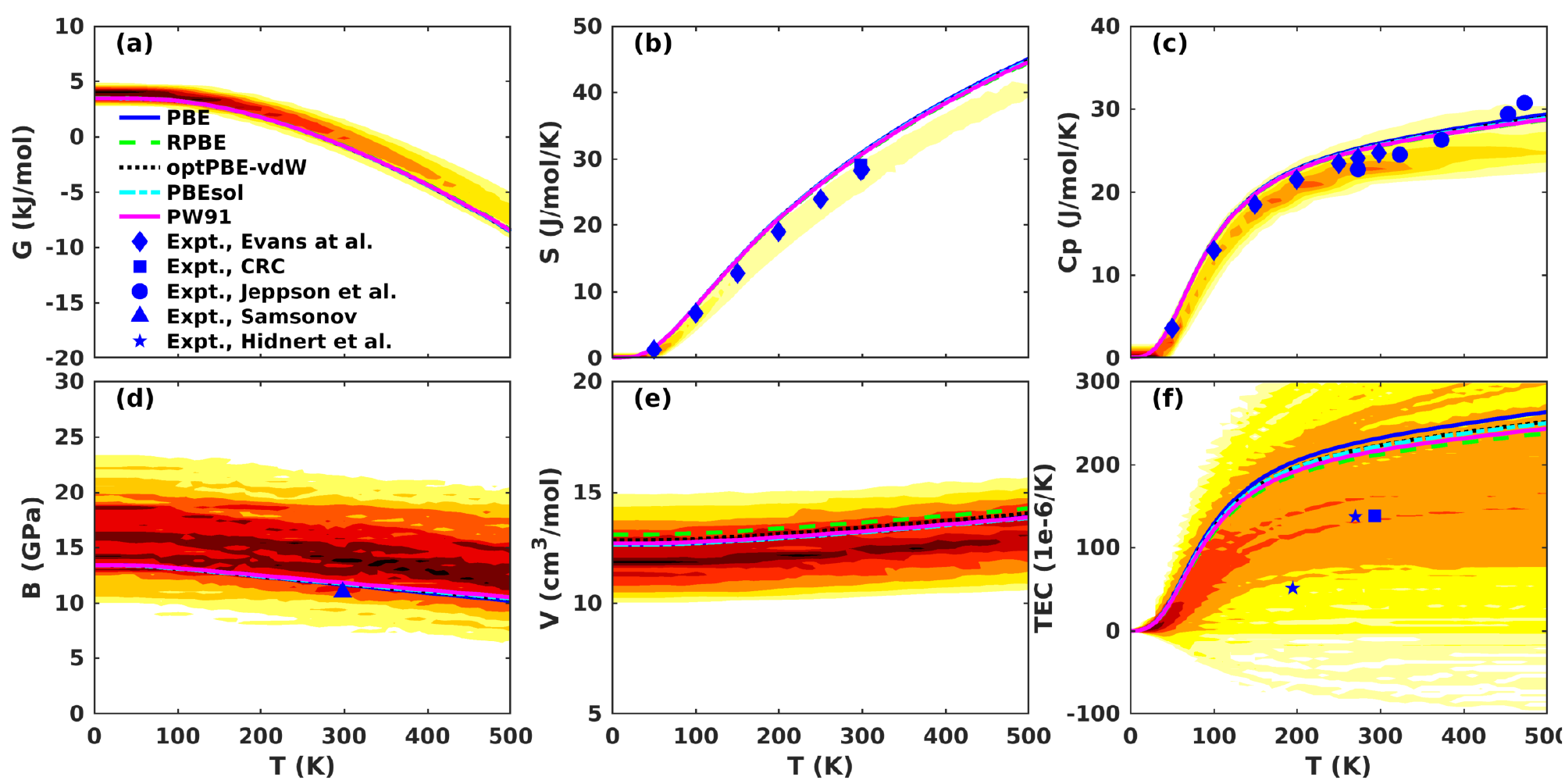}
\centering
\caption{Thermodynamic properties of bcc Li: (a) Gibbs energy (b) entropy (c) isobaric heat capacity (d) bulk modulus (e) volume and (f) volumetric thermal expansion coefficient from DFT Debye calculations compared with experimental data in the literature \cite{evans1955thermodynamic,alma991000880309704436,osti_6885395,samsonov2012handbook,american1963american}. The colormap represents the PDF of each property calculated by BEEF. The natural logarithm of PDF is presented in (f) for better view.}
\label{fig:Li}
\end{figure}

\begin{figure}[H]
\includegraphics[width=\textwidth]{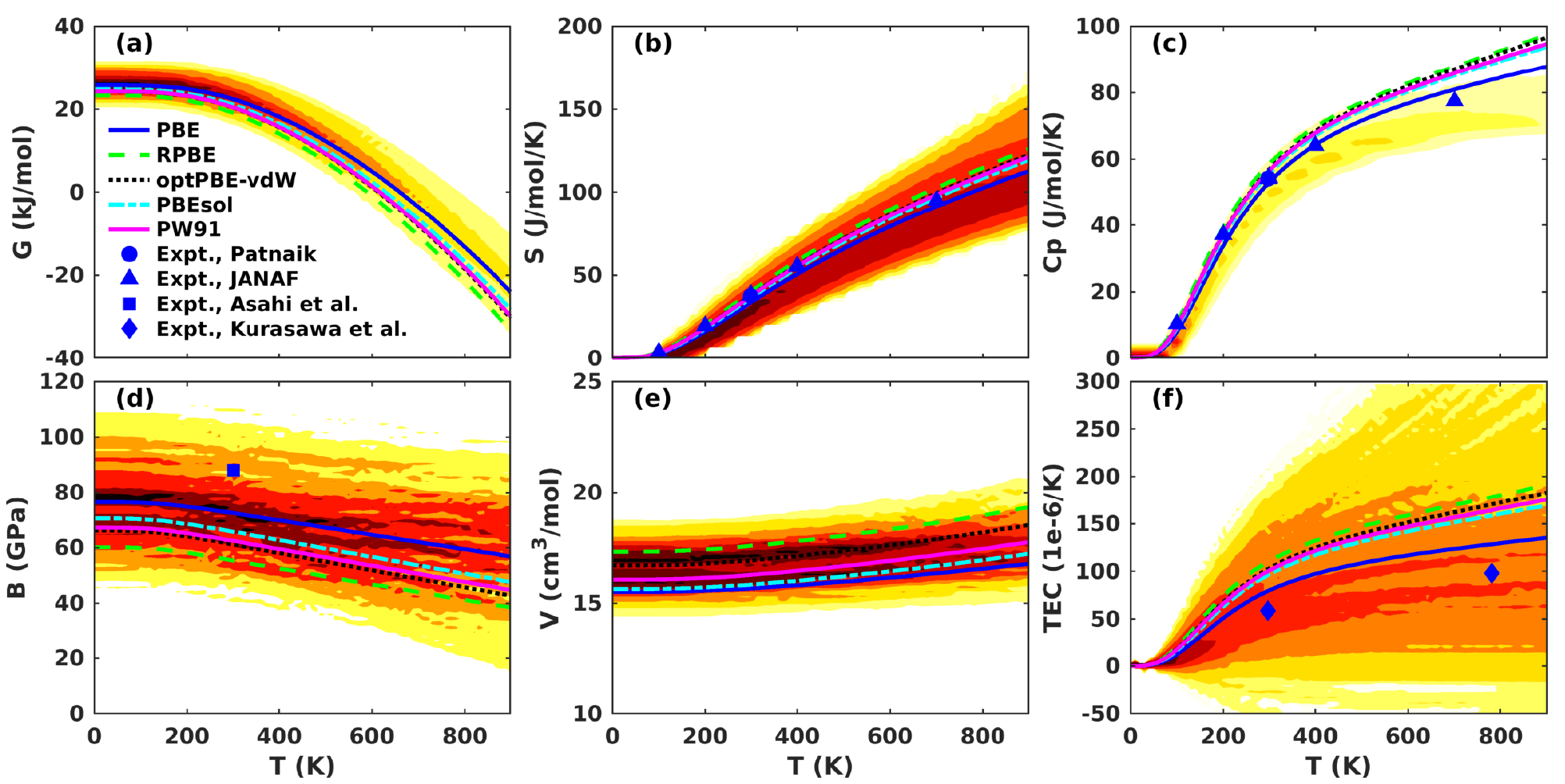}
\centering
\caption{Thermodynamic properties of anti-fluorite \ce{Li2O}: (a) Gibbs energy (b) entropy (c) isobaric heat capacity (d) bulk modulus (e) volume and (f) volumetric thermal expansion coefficient from DFT Debye calculations compared with experimental data in the literature \cite{patnaik2002handbook,stull1971janaf,Asahi_2014,KURASAWA1982334}. The colormap represents the PDF of each property calculated by BEEF. The natural logarithm of PDF is presented in (b) and (f) for better view.}
\label{fig:Li2O}
\end{figure}

Eight materials are chosen to study the effect of DFT functionals on the prediction of thermodynamic properties within the Debye model. These materials belong to seven crystal structures: fcc (Al and Ca), bcc (Li), hcp (Mg), rocksalt (NiO), zincblende (GaAs), anti-fluorite (\ce{Li2O}) and corundum (\ce{Al2O3}), covering a wide spectrum of material structures in metals, semiconductors and insulators. The thermodynamic properties of the chosen materials are shown in \crefrange{fig:Mg}{fig:Li2O}, where the probability distribution function (PDF) of each property calculated by BEEF is compared with the values obtained from five functionals (PBE \cite{PhysRevLett.77.3865}, RPBE \cite{PhysRevB.59.7413}, optPBE-vdW \cite{Klime__2009}, PBEsol \cite{PhysRevLett.100.136406} and PW91 \cite{PhysRevB.54.16533}), as well as available experimental data. The values for the predictions from all of these functionals as well as the statics of the ensemble of BEEF predictions can be seen in Table \ref{tab:stat1} for 300K, with another table of statistics at 900K available in the Supporting Information.

Here we present a more detailed discussion of the prediction results and estimated uncertainty the materials tested. To aid in this discussion we include detailed plots for four materials, with the remaining plots for all other materials in the Supporting Information. We find a good agreement between the computational predictions and experiment for most properties. In particular, we find very good agreement in the predictions of entropy, S, and specific heat, C$_P$ except as the temperature approached the melting point of the material. As seen in Figure \ref{fig:Li} we see for Li, with a melting point of around 450K, a rise in the experimental heat capacity near this temperature that cannot be captured by the Debye model. The largest uncertainties, as determined by the BEEF ensemble, as well as the largest disagreement between the functionals tested and experiment, are present in the predictions of bulk modulus (B) and thermal expansion coefficient (TEC). In the case of GaAs and Li (Figure \ref{fig:Li}) respectively, we see a systematic over estimation of the thermal expansion and for all materials. Additionally, over all materials the thermal expansion has the largest spread of any property, predicting the largest relative uncertainty in this property. The shape of this distribution at each temperature, however has a thinner tail and sharper middle peak than that of a normal distribution as seen in it's large positive excess Kurtosis. This means that although there are a lot of outliers very far away from the mean, the majority of prediction are relatively close to the mean.

We see a similar trend for the underestimation of the bulk modulus of NiO, GaAs, and \ce{Al2O3} (Figure \ref{fig:Al2O3}, while there is relatively good agreement for the bulk modulus of Al, Li, Mg, and Ca. These errors are likely due completely to the GGA level DFT overbinding the strained and compressed volumes and underbinding the equilibrium volumes as even at 0K, when there is no influence of the Debye model approximation on the predictions, the uncertainty remains large and constant when looking at the coefficient of variation. Additionally, unlike in the case of thermal expansion coefficient, the distributions for the bulk modulus have much fatter tails and a broader peak, as described by the negative excess Kurtosis. 

Additionally, we find that the ensemble of BEEF predictions captures the predictions of the other GGA level functionals. In this way, we see that BEEF is providing a reasonable sampling of GGA-level DFT space through the non-self consistent evaluations of a distribution of exchange correlation functionals. In most cases, the distributions for Gibbs energy (and as a result Zero point energy), entropy and and heat capacity are small, especially in the case of  Al, NiO, GaAs, and \ce{Al2O3}. As mentioned previously, the computational predictions match the experiment and we find BEEF to have correctly predicted the relatively low uncertainty in these predictions. 

\subsection{Distribution Shapes}
The way that the underlying distribution of the energetic predictions generated from DFT with BEEF is transform through the Debye model is complex. Due to steps such as fitting of an equations of state where the parameters of the fit are then passed on through integral and derivative equations, the analytic transformation of the data is hard to describe. To solve this, we fit the final distributions to a collection of model probability distribution functions in an attempt to find the true statistical nature of the data generation process for that property. 

The distributions of various properties for NiO at 300K have been plotted. Similar plots for other materials and temperatures are available in the Supporting Information to understand the variation of the distributions over materials and temperature. The collection of model distributions used to fit the data consists of normal, generalized normal, skewed normal, and other transformations of the normal distribution. It is likely that the true data generation process for many property's probability distribution is related to a normal distribution as the spread of the BEEF-vdW energetic predictions for a single volume is a normal distribution. Additionally to test the possibility of a less normal-like distribution, a generalized Gamma and a Gamma-like Weibull distribution were also tested, as well as a Gumbel and a generalized extreme value distribution to attempt to model the volume variable as it is the extreme value (minimium) of the equation of stat curve fit. For each property, a Cramer von Mises goodness of fit test\cite{anderson1952asymptotic} was performed to test the hypothesis that the model distribution and the sample distribution are identical. We then compared the two model distributions with the best Cramer von Mises statistic with a Vuong likeness test, that compares the Kullback–Leibler divergence from the data for each model distribution. The goodness of fit metric allows us to test if there is statistical evidence that the model distribution is not the true data generation process or not. On the other hand, the Vuong closeness tests, examines how much closer the chosen model is to the true data generation process than another possible model. For the majority of BEEF ensemble property data fit, we find good statistical agreement with our fits.

\begin{figure}[H]
\includegraphics[width=\textwidth]{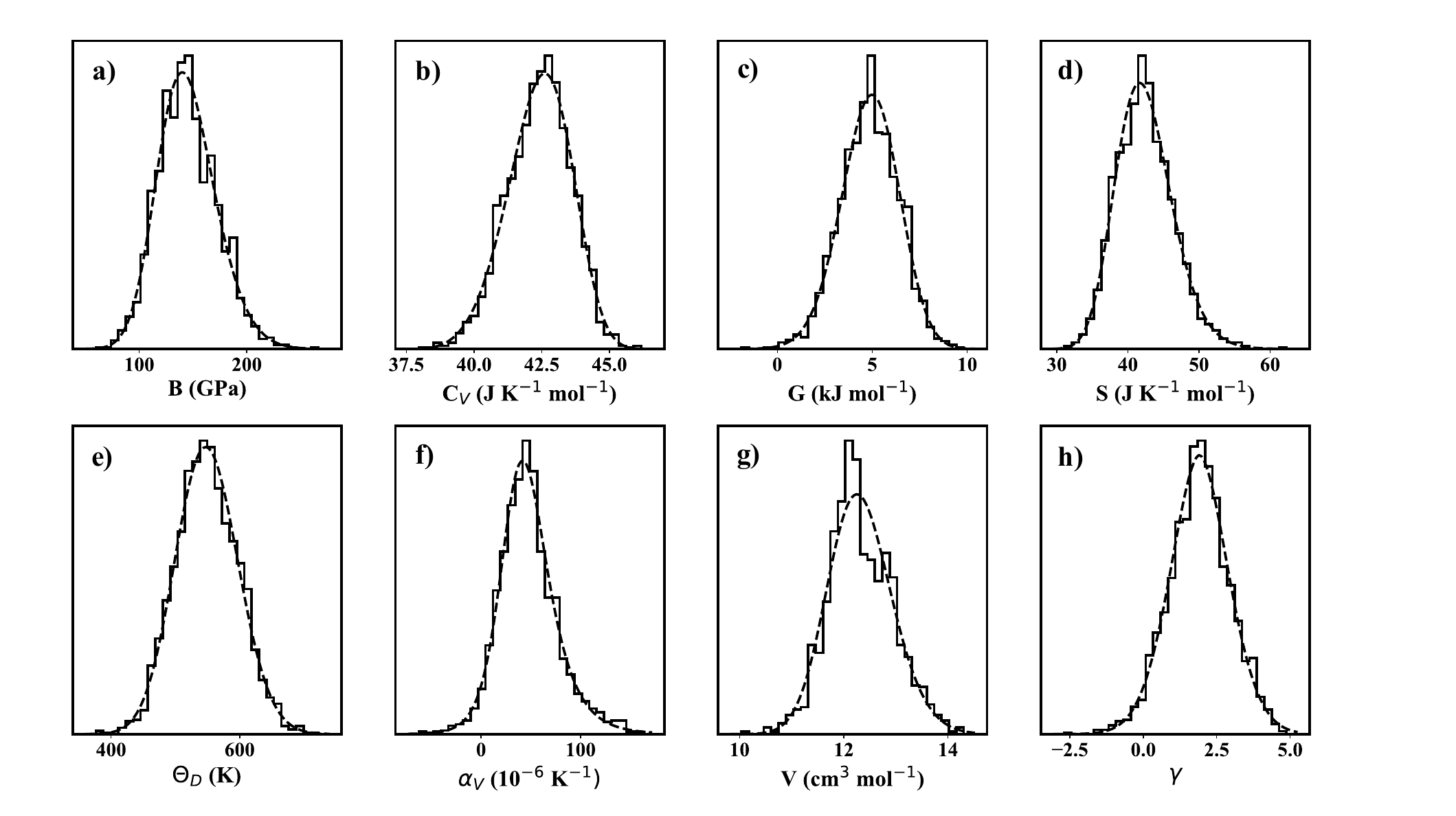}
\centering
\caption{The distribution of thermodynamic properties (a) bulk modulus, (b) isochoric heat capacity, (c) Gibbs free energy, (d) entropy, (e) Debye temperature, (f) volumetric thermal expansion coefficient $\alpha_V$, (g) volume at zero pressure, and (h) the Gr{\"u}neisen parameter $\gamma$ of rocksalt NiO at 300.}
\label{fig:NiO-300}
\end{figure}

The results of these statistical fittings and tests is shown for NiO at 300K in Table \ref{tab:NiO_300fit}. In this table, the p-value of the Cramer von Mises test is shown. This is the chance of error when rejecting the null hypothesis that the data we have was drawn from the model distribution. If this is above 0.05 we say there is no statistically significant evidence that the model distribution and real distribution of the data are different. The p-value result of the Vuong test for comparing the two best distributions is also shown. This is is the chance of error when assuming the chosen distribution is better than the second best distribution at describing the data. If this is below 0.05, we say there is enough statistical evidence to conclude the chosen model is closer to the true data generation process than the second best model. Within the table, we see some p-values of the Vuong test that are above 0.5, meaning the results of the Vuong test are leaning towards the second best model from the goodness of fit test, is better at describing the data. In no case, however, where there enough evidence from the Vuong test to reject the results of the goodness of fit at the level of 0.05 significance and therefore we always kept the fit with the best fit. For most properties, a distribution was found with an extremely high confidence of fit, even if the Vuong test is suggesting an equally good fit is available with another model. This is likely due to the similarities in the set of models chosen within the work. In the case of volume, however, the result of the Cramer von Mises tests would suggest you reject at the 0.05 level that our model distribution describes the data, but that the model is conclusively better than all other models. This is likely due to a slight bi-modality in the distribution of volume predictions as seen in Figure \ref{fig:NiO-300}. This is probably due to the error in the equation of state fits and the inability for this type of UQ to account for differing qualities of fit.

\begin{table}
    \centering
	\caption{\label{tab:NiO_300fit} The fit for the properties of NiO at 300K with the result of the Cramer von Mises (CM) goodness of fit test and Vuong Test. Here $\phi(x)$ is the normal pdf and $\Phi(x)$ is the normal cdf.}

		\begin{tabular}{ccccc}
			
			Property & Form &  & CM & Vuong Test\\ \hline
			\\
			$B_T$& $f(y,c)=2\phi(y)\Phi(cy)$  & $c=1.616$ & 0.85 & 0.58\\ 
			& & $y=\dfrac{x-119.7}{37.37}$ & &\\
			\\
			$C_V$&$f(y,c)=cx^{c-1}\exp{-x^c}$  & $c=4.966$ & 0.96 &0.645\\
			& & $y=\dfrac{x-37.28}{5.578}$& &\\
			\\
			$G$  &$f(y,a,c)=\dfrac{|c|x^{ca-1}\exp{-x^c}}{\Gamma(a)}$& $a=9.611$ &0.81 & 0.38\\ 
			& & $c=6.039$& &\\
			& & $y=\dfrac{x+23.23}{19.45}$& &\\
			\\
			$S$  & $f(y,a,c)=\dfrac{c}{\sqrt{y^2+1}} \phi(a+c\sinh^{-1}(y))$ & $a=-7.155$ & 0.83 &  0.43\\
			& & $c=6.204$& &\\
			& & $y=\dfrac{x-22.18}{14.01}$& &\\
			\\
			$\Theta_D$&$f(y,c)=c\phi(y)\Phi(-y)^{c-1}$&$c=0.653$ & 0.88 & 0.23\\
			& & $y=\dfrac{x-531.81}{42.54}$& &\\
			\\
			$\alpha_V$& $f(y,a,c)=\dfrac{c}{\sqrt{y^2+1}} \phi(a+c\sinh^{-1}(y))$& $a=-0.5960$  &0.79& 3.1$\times 10^{-3}$\\
			& & $c=1.792$& &\\
			& & $y=\dfrac{x-30.60}{43.33}$& &\\
			\\
			$V$    & $f(y,c)=2\phi(y)\Phi(cy)$ & $c=1.278$ & 0.043 & 0.013\\
			& & $y=\dfrac{x+11.83}{0.7954}$& &\\
			\\
			$\gamma$&$f(x,c)=\dfrac{c}{2\Gamma(1/c)}\exp{-|x|^c}$ &$c=1.811$&0.86&0.16\\
			& & $y=\dfrac{x-1.910}{1.388}$& &\\
			  
		\end{tabular}

\end{table}

\section{Conclusion}

Using the Debye model and density functional theory for the prediction of thermodynamic properties, the effect of uncertainty on these predictions due to the the underlying uncertainty of the DFT predictions has been explored. Based on the calculation of eight materials including metals, semiconductors and insulators in seven structures, it is found that the predictions of GGA-level functionals (PBE, RPBE, optPBE-vdW, PBEsol and PW91) are bounded by the estimates from the BEEF ensemble distribution. The uncertainty of temperature-dependent part of Gibbs energy predicted by BEEF increases with temperature and can be quite significant for determining phase boundaries. Finally, we test a wide variety of probability distributions that describe the finite temperature distribution and find that most properties are well described by normal or transformed normal distributions.   However, the spread of volume at a given temperature does not appear to be drawn from a single distribution.  The present work consists of a preliminary attempt to quantify the uncertainty of thermodynamics from first-principles. Further extensions may include adopting more complicated models such as calculations of phonon spectrums, as well as uncertainty associated with different levels of approximations on the models.  In order to accelerate adoption of these methods, we have open-sourced the code-base under the name, Depye.

\begin{acknowledgement}

This work was also partially supported by the Assistant Secretary for Energy Efficiency and Renewable Energy, Office of Vehicle Technologies of the US Department of Energy (DOE) through the Advanced Battery Materials Research (BMR) Program under contract no. DE-EE0007810.

\end{acknowledgement}

\begin{suppinfo}
Debye Model Analysis for other studied materials.
\end{suppinfo}

\bibliography{cite}





\includepdf[pages=-]{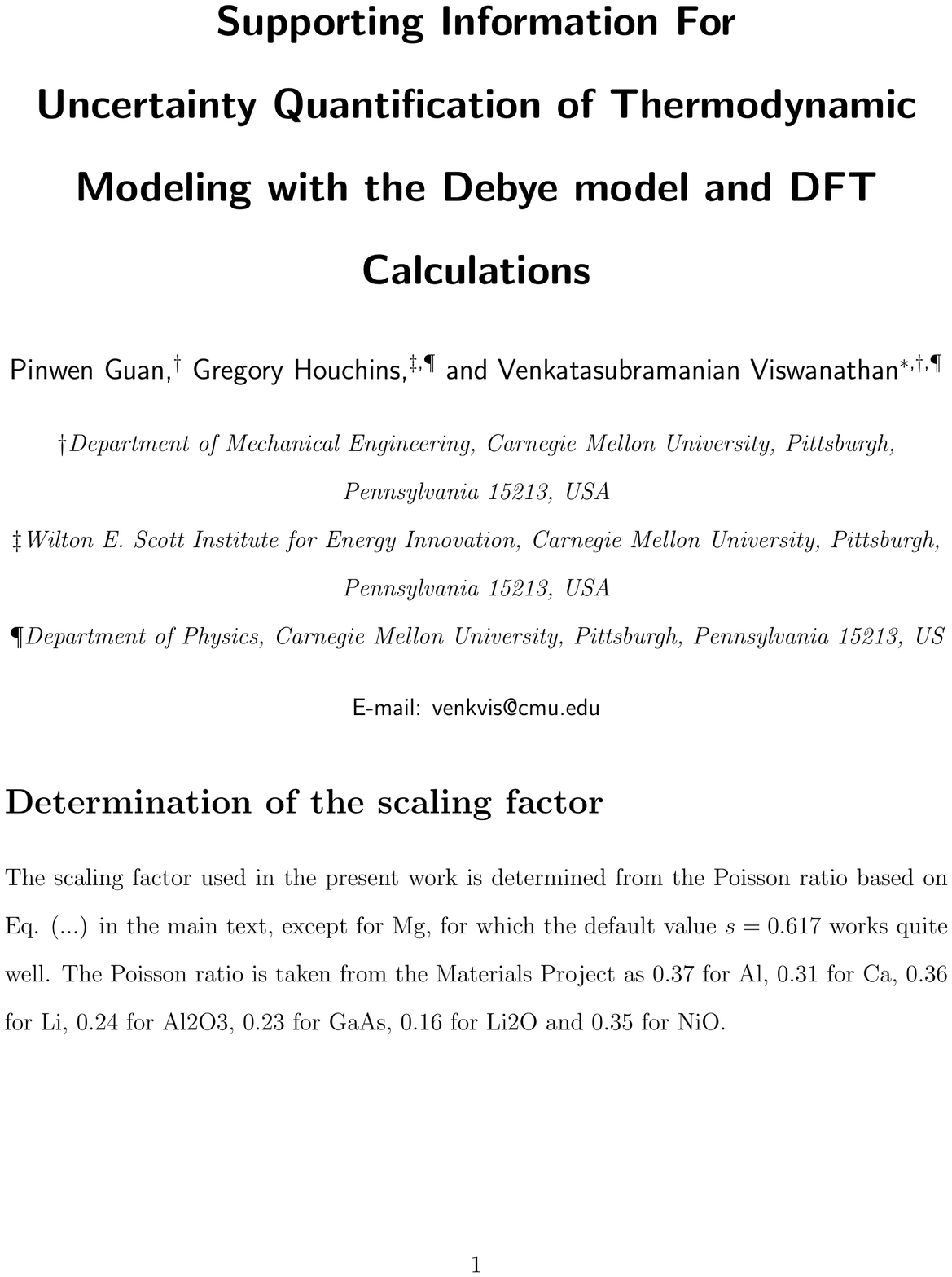}

\end{document}